\documentstyle[preprint,aps]{revtex}
\begin{document}

\title
{Analysis of Photoreaction in the Delta Energy Region\\
by the Quantum Molecular Dynamics Approach}

\author{
Tomoyuki Maruyama$^{(1)}$,
Koji Niita$^{(1,2)}$,\\ 
Satoshi Chiba$^{(1,3)}$, 
Toshiki Maruyama$^{(1)}$
and 
Akira Iwamoto$^{(1)}$}

\address{
$^{1}$Advanced Science Research Center, 
Japan Atomic Energy Research Institute, \\
Tokai-mura, Naka-gun, Ibaraki-ken 319-11, Japan\\
$^{2}$Research Organization for Information Science 
and Technology \\
Tokai-mura, Naka-gun, Ibaraki-ken 319-11, Japan \\
$^{3}$Nuclear Data Center, 
Japan Atomic Energy Research Institute, \\
Tokai-mura, Naka-gun, Ibaraki-ken 319-11, Japan\\
}

\maketitle


\begin{abstract}
We study the photoreaction in the delta energy region using the QMD approach.
The proton and pion cross-sections are calculated and compared with 
experimental data.
Through this work we examine the multistep contributions in the cross-sections
and the {$\pi$-$\Delta$} dynamics.
\end{abstract}



\newpage

\section{Introduction}\label{introduction}

$\Delta$ properties in medium has been investigated both in experiments
and in theories.
J. Chiba et al. \cite{chibaj} suggested experimentally 
that the {$\Delta$} mass becomes lower in nuclear medium than that 
in the free space.
Horikawa et al. \cite{horikawa} gave that the depth of 
the {$\Delta$}-potential is about 30 MeV in normal nuclei.
On the other hand the recent total photoabsorption experiments \cite{totab} 
showed the broadening of the {$\Delta$} peak but no shift of the peak-position.
These works indicated that {$\Delta$} properties are modified
in nuclear medium, but this medium correction has not been understood definitely.

Experiments with photon in the {$\Delta$} energy region are expected to 
has some advantages, compared with other experiments, to get 
information of {$\Delta$} properties in medium.
First the photon can directly produce {$\Delta$} in the bulk region 
since the electromagnetic interaction is weak, 
while the proton- and pion-induced reactions produce it mainly in the surface region.
Second the produced {$\Delta$} does not have so large momentum because a value 
of momentum transfer is same as that of energy transfer for photoreaction.
In the proton-induced reaction the momentum transfer is much larger than the
energy transfer, so that the produced {$\Delta$} must have a large momentum,
and in-medium properties do not strongly affect observables.

The inclusive experimental data of photoabsorption is
not sufficient to identify the elementary absorption process uniquely.
We need to observe a outgoing nucleon and/or a pion to investigate in-medium {$\Delta$}
properties.
In fact TAGX collaboration at INS \cite{TAGX,Emura} showed in experiments 
of photoreaction with the $^{3,4}$He target that one must 
observe several outgoing particles coincidentally 
so as to identify the pure {$\Delta$} channel.
These particles, however, interact with other nucleons before they escape the nucleons
and have lost information
of the photoabsorption process at the beginning.
Thus we have to analyze the multistep collisional processes
after the photoabsorption. 

Recently we developed a framework of QMD \cite{Aich} plus statistical decay model 
(SDM) \cite{niita95}, 
and applied systematically this QMD $+$ SDM to nucleon- (N-) induced reactions.
It was shown \cite{niita95}
that this framework could reproduce quite well
the measured double-differential cross sections of 
(N,xN') type reactions from 100 MeV to 3 GeV 
incident energies in a systematic way. 
In the subsequent papers \cite{chadwick95,chiba96}, we gave detailed
analysis of the pre-equilibrium (p,xp') and (p,xn) reactions
in terms of the QMD in the energy region of 100 to 200 MeV.  
In these analysis, a single set of parameters was used, and no  
readjustment was attempted.  

The reaction process after the initial photoabsorption 
is almost the same as the preequibrilium process
of nucleon-induced reactions.  
It should then be natural to apply the QMD $+$ SDM approach
to the analysis of the electron scattering and photoreaction.
Of course we have the other methods to calculate the multistep
contributions such as the PICA code \cite{PICA}, a Monte-Carlo (MC) model 
\cite{carrasco} and BUU \cite{hombach,effenberger,BUU4}.
The PICA does not have a {$\Delta$}-degree of freedom explicitly, 
and it is not very useful to study the $\pi$-$\Delta$ dynamics.
The MC calculation is performed only in the momentum-space, and cannot describe
the refraction from the mean-field.
The BUU approach has succeeded to describe the particle production in the
intermediate energy heavy-ion collisions.
However this approach can describe only the one-body dynamics,
and then it cannot distinguish single nucleons from clusters and
cannot calculate coincident observables.
Anyway none of these models can treat the nucleon-induced reactions, 
heavy-ion collisions
and photoreactions (electron scattering) in the uniform way.
The ability of the uniform description is one of the strongest advantages 
of the QMD approach.

We then carry out an analysis of photoreactions with the same formula and 
the same set of parameters as the previous works of the nucleon-induced 
reactions\cite{niita95,chadwick95,chiba96}.
This enables us to check and improve further the elementary collisional 
processes included in the QMD approach from another point of view.

In this paper, we focus only on the photoreaction in the energy region
above the {$\Delta$} threshold.
In the next section, a brief explanation of the QMD plus
SDM approach is given.  The comparison of the calculation with
the experimental data and discussions on
the reaction mechanisms are given in section \ref{results}. 
Summary of this work is given in section \ref{summary}.

\section{Brief explanation of The Quantum Molecular Dynamics}
\label{brief}

\subsection{Equation of motion}
We start from representing each nucleon (denoted by a 
subscript {\it i}) by a Gaussian wave packet 
in both the coordinate and momentum spaces.
The total wave function is assumed to be a direct 
product of these wave functions.
Thus the one-body distribution function is obtained 
by the Wigner transformation of the wave function,
\begin{equation}
f({\bf r}, {\bf p}) = \sum_i f_i({\bf r}, {\bf p})
= \sum_i 8 \cdot \exp \left[ -\frac{({\bf r}-{\bf 
R}_i)^2}{2L} 
-\frac{2L(\bf{p}-\bf{P}_i)^2}{ \hbar ^2} \right]
\label{eq:phase-space}
\end{equation}
where ${\it L}$ is a parameter representing the spatial spread of 
a wave packet, 
${\bf R}_{\it i}$  and ${\bf P}_i$ corresponding to
the centers of a wave packet in the coordinate and momentum spaces, 
respectively.
The equation of
motion of ${\bf R}_i$  and 
${\bf P}_i$ is given, on the basis of the time-dependent variational 
principle, by the
Newtonian equation:
\begin{equation}
\dot{\bf R}_i=\frac{\partial H}{\partial {\bf P}_i}, ~~~~~
\dot{\bf P}_i=-\frac{\partial H}{\partial {\bf R}_i}, 
\label{eq:eos}
\end{equation}
and the stochastic N-N collision term \cite{niita95}.
We have adopted the Hamiltonian 
$\it{H}$ consisting of the relativistic kinetic and mass energies
and the Skyrme-type effective N-N 
interaction\cite{skyrme} plus Coulomb and symmetry energy terms:  
\begin{eqnarray}
H~=~& &\sum_{i}\sqrt{m^2_i~+~\bf{P}^2_i} \nonumber \\
& &{} +\frac{1}{2} \frac{A}{\rho_0} \sum_{i}<\rho_i>
+\frac{1}{1+\tau} \frac{B}{\rho_0^\tau} \sum_{i}<\rho_i>^\tau 
\nonumber \\
& &{} +\frac{1}{2}\sum_{i,j(\neq i)}\frac{e_i e_j} {\mid 
\bf{R}_i-\bf{R}_j \mid}
   {\rm erf} \left( \mid \bf{R}_i-\bf{R}_j \mid /\sqrt{4 {\it L}} 
\right) \nonumber \\
& &{} +\frac{C_s}{2 \rho_0} \sum_{i,j(\neq i)} c_i c_j \rho_{ij},
\label{eq:hamiltonian}
\end{eqnarray}
where "erf" denotes the error function, the $e_i$ is the charge of the $i$-th
particle, and the $c_i$ is 1 for proton, -1 for neutron and 0 for other
particles.
With the definition
\begin{eqnarray}
\rho_i({\bf r})&\equiv &\int \frac{d{\bf p}}{(2 \pi \hbar)^3}f_i({\bf 
r},{\bf p})\nonumber \\
{}&=& (2 \pi L)^{-3/2} \exp \left[-({\bf r}-{\bf R}_i)^2/2L \right] ,
\label{eq:rhoi}
\end{eqnarray}
the other symbols in eq.(\ref{eq:hamiltonian}) are given as:
\begin{eqnarray}
<\rho_i>&\equiv & \sum_{j (\ne i)} \rho_{ij} \equiv \sum_{j (\ne i)}
  \int d {\bf r} \rho_i({\bf r}) \cdot \rho_j({\bf r})\nonumber \\
 {}&=& \sum_{j (\ne i)}(4 \pi L)^{-3/2} \exp \left[-({\bf R}_i-{\bf 
R}_j)^2/4L \right].
\label{eq:hoho}
\end{eqnarray}
The symmetry energy coefficient $C_s$ is taken to be 25 MeV.  The four
 remaining parameters, 
the saturation density  $\rho_0$, Skyrme parameters $A$, $B$ and 
$\tau$
are chosen to be 0.168 fm$^{-3}$, $-124$ MeV, 70.5 MeV and  4/3, 
respectively. 
These values give the binding energy/nucleon of 16 MeV at the saturation 
density $\rho_0$ and the compressibility of 237.7 MeV (soft EOS) 
for nuclear matter limit.  
The only arbitrary parameter in QMD, i.e., the width parameter {\it 
L}, is fixed to be 2 
fm$^2$ to give stable ground state of target nuclei in a wide mass 
range.
These values and  the details of the other description are just the same as 
those according to our previous paper\cite{niita95}.

As for the isobar resonances such as $\Delta$ and $N^*$, we use
the same interactions as nucleons though the symmetry force
dose not work for them.
At each collision process we satisfy the energy conservation by 
varying slightly an absolute value of relative
momentum between colliding two particles.

\subsection{The initial state of the photoabsorption}

The QMD calculation is started at the moment 
when the photon is absorbed by the target nucleus.
As an initial state of the simulation,
we have to assume the photoabsorption channels.
For 200 - 400 MeV/c  incident photon momenta,
the one-pion production process is dominant in the photoabsorption.
We introduce the following three channels 
for the photoabsorption, i.e.,
\begin{eqnarray}
\gamma + N \rightarrow & \Delta, ~~~ &  ~~(C1)
\nonumber \\
\gamma + N \rightarrow & N^*, ~~~ &  ~~(C2)
\nonumber \\
\gamma + N  \rightarrow & N + \pi. ~ &  ~~(C3)
\end{eqnarray}
The nuclear resonances $\Delta$ and $N^{*}(1440)$ 
in the above equations
can decay into nucleon and pion
according to the decay width and the isospin selection.
In addition, the $N^*$ can also decay into $\Delta$ and $\pi$
as in our prescription \cite{niita95}, which means that
the (C2) channel includes implicitly two-pion production process.
The channel of the pure ${\pi}N$-pair production (C3)
is adopted only for the charged pion
and the isospin symmetry is assumed for the cross section, i.e.,
$\sigma(\gamma + p) = \sigma(\gamma + n)$.

The cross sections of the above three channels are determined 
to reproduce 
the experimental one-pion production data of 
$\gamma + p \rightarrow N + \pi$ \cite{eldata,eldata2}.
The results are given in Fig. \ref{gnpi}.
The upper part (a) of Fig. \ref{gnpi} denotes
the cross-section of $\gamma + p \rightarrow p + \pi^0$ and  
the lower part (b) denotes that of 
$\gamma + p \rightarrow n + \pi^{+}$.
The long dashed, dashed and thin solid lines indicate contributions
of the $\Delta$ resonance (C1), $N^{*}(1440)$ (C2) 
and the pure ${\pi}N$-pair production (Born term) (C3), respectively.
For the future discussion, we define the alternative channel (C4)
that the all photon is absorbed through the pure ${\pi}N$-pair
with the same amount of
the cross-section as the sum of $C1 + C2 + C3$ channels. 

In each event, a nucleon which absorbs the photon is selected randomly.
The photoabsorption channel is also randomly chosen
according to the rate of each cross section. 
We assume that the angular distribution of the pion emission is isotropic
in the center of mass system of the pion and the nucleon.

\subsection{Decomposition into step-wise contribution in multistep 
reactions}

For the later discussion of the multistep reaction, 
we define here 
the step number $s$ which indicates 
the number of collisions responsible for emission of a particle
in the QMD calculation as following.  
First we assign the step number zero to each nucleon in 
the target nucleus.  
After a nucleon absorbs the incident photon and becomes a resonance, 
we set the step number of the resonance to be one.
For the case of the pure ${\pi}N$-pair creation, 
the step numbers of both $\pi$ and $N$ are one.
The rule of the change of the step number for 
each nucleon is that, if two nucleons
{\it i} and {\it j} having step numbers $s_i$ and $s_j$ make a 
collision, the step numbers 
of both particles are modified to be $s_i + s_j + 1$.  
If a pion $i$ and nucleon $j$ becomes a resonance,
the step number of the resonance is also  $s_i + s_j + 1$.  
When a resonance $i$ decays into a pion and a nucleon, 
on the other hand, 
both the step numbers of the pion and the nucleon are set to be $s_i$;
namely this process does not change the step number.
We prohibit successive collisions with the same partner
and the collisions between two nucleons with 0-step number.
We attach n-step contribution to the ($\gamma, N$) or ($\gamma, \pi$) 
reactions when a nucleon or a pion emitted from the target nucleus 
has a step number $n$.

\subsection{Calculation of the Cross-Section}
\label{cross-section}

In the simulation,
the double-differential 
cross-section of the emitted particle is calculated as 

\begin{equation}
\frac{d^{2}\sigma}{d E \ d \Omega}~=~\frac{1}{N_{event}} 
\sum_{i} \sigma^{T}_{\gamma}(i)  M(E,~\Omega,~i) 
\label{eq:cross-section}
\end{equation} 
where $i$ indicates the event number,  $N_{event}$ is the total number of 
the events, $\sigma^{T}_{\gamma}(i)$ shows the total photoabsorption 
cross-section in the $i$-th event, and 
$M(E,~\Omega,~i)$ denotes the multiplicity of the particle 
under interest emitted in the unit energy-angular interval around
$E$ and $\Omega$ for the $i$-th event.
    
Typically, 400000 events were generated to get a reasonable statistics 
in the 
step-wise double-differential cross-section.  In the calculation, the 
parameters have been fixed to the same values as in Ref. \cite{niita95} 
without any adjustment.

\section{Results and Discussion}
\label{results}

In Fig.~\ref{gC375}, we show our results of proton (left-hand-side) and 
$\pi^0$ (right-hand-side) double differential cross sections 
at $\theta = 30^{\circ}, 60^{\circ}$ and $90^{\circ}$
from $\gamma (375 MeV/c) + {\rm C}$  reaction.
The experimental data at $\theta = 30^{\circ}$ for proton,
which are denoted by the full circles with 
error bars, are taken from Ref. \cite{Kanazawa}.
In these figures, we decompose the total cross sections into the step-wise 
contributions.
The total results of the QMD+SDM simulation are shown by the thick lines.
In the same figures, we draw 
the individual contributions from the 1-step process (long-dashed lines),
2-step (dashed lines), 3-step (thin lines) and SDM process (dotted lines),
respectively.
The contribution from SDM process is shown only for proton.

First we look at the contribution from the 1-step process.
The peak of the 1-step process, which is usually called
quasi-free (QF) peak, is clearly separated from the multistep 
contribution in the forward angle ($\theta = 30^{\circ}$),
while it overlaps to other contributions at larger angles,
particularly for proton.

For the more detailed analysis of the step-wise contributions,
we plot again the the results of the proton cross-section 
at $\theta = 30^{\circ}$ from $\gamma (375 MeV/c) + {\rm C}$  reactions
in Fig. \ref{gC375ch}.
The upper column (Fig. \ref{gC375ch}a) is just the same 
as that in Fig.\ref{gC375}. 

We can see in this figure that the position of the calculated QF peak 
almost coincides with the experimental peak position,
though the absolute value of this peak overestimates the data. 
This peak comes mainly from 
the 1-step process, namely from the elementary process
of $\gamma + N \rightarrow \pi + N$.

There are two other peaks in our results.
The peak in the lower momentum region shows the contribution of SDM,
the evaporation from the residual excited nuclei.
The higher momentum peak, on the other hand, mainly comes from 2-step 
process.
There is a dip region between this peak and the QF peak,
which are not observed experimentally.

In order to understand the meaning of the higher momentum peak,
we also decompose contributions to the proton cross-section of the events 
with zero pion and with one pion at the final state in Fig. \ref{gC375ch}b.
This figure clearly shows that the cross-section around the higher momentum 
peak comes only from 
the zero pion events .
>From this analysis we can easily know that the cross-section around 
the higher peak comes from the following 2-step process:
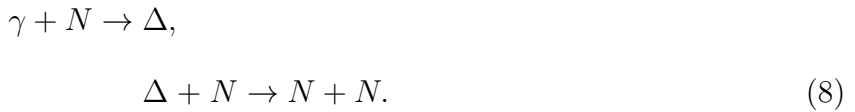
\begin{eqnarray}
\gamma + N   \rightarrow & \Delta & , 
\nonumber \\
& \Delta & \; + \; N   \rightarrow   N + N .
\end{eqnarray}
This 2-step process is effectively equivalent to 
the two-nucleons ($2N$) photoabsorption process.

In Fig. \ref{gC375ch}, we give also the result (chain-dotted line)
of QMD calculation including only the pure ${\pi}N$ pair as the
initial channel (C4 channel mentioned before).
It is seen that this result does not show the higher peak of $2N$ 
photoabsorption any more. 
Hombach et al. have commented in Ref. \cite{hombach} that these two channels do not
make large difference in $\pi$-productions.
At least proton spectra, however, there is important diffrence between the two initial
channels.

In Fig. \ref{g390p}, we show the results of proton energy-spectrum from 
the photoreaction at the photon-momentum $q = 390 MeV/c$ with the target
$^{12}$C (Fig. \ref{g390p} (a) ) and $^{48}$Ti (Fig. \ref{g390p} (b) ).
We draw the full results of the QMD+SDM simulation and 
the individual contributions from the 1-, 2- and 3- step processes
as in the previous figures.
The experimental data are taken from Ref. \cite{Arends91}. 
In order to compare each contribution in detail, we plot the results with the
logarithmic scale.
In this reaction the 1-step and 2-step processes make almost same contributions
to the cross section at low proton energy, 
and  the QF peak cannot be seen clearly.
Our QMD+SDM results underestimate experimental data around
the QF peak and also $2N$-photoabsorption energy regions.
The same behavior was seen in the results of PICA for this reaction 
\cite{Arends91}.

In the above two comparisons with the experimental data,
it is seen that around the QF peak energy region
our calculations overestimate the experimental data 
at $\theta = 30^{\circ}$ in Fig. \ref{gC375ch}, but underestimate 
at $\theta = 52^{\circ}$ in Fig. \ref{g390p}.
It has been already known from the analysis of the proton-induced reaction
\cite{niita95,chadwick95,chiba96}
that the QF contribution strongly depends on the detailed of the
elementary process.
In this energy region, $q \approx 380 MeV/c$, the angular distribution 
of emitted pion has a sideward peak in the CM system of the ${\pi}N$-pair
\cite{eldata}.
We have not fit the angular distribution of the pion photoproduction
cross-section to this experimental data in the elementary process,
and this might be the reason for the disagreement near the QF peak.

For the higher momentum region, on the other hand, our results seem 
to agree with the data and imply that the multistep process, mainly 
the 2-step process, can explain the data
for the photon energy above the pion threshold.

Next we explore the origin of the dip region between the above two momentum 
region.
The experimental analysis \cite{Emura} with $^{3}$He target
indicated that the three-nucleons ($3N$) photoabsorption process 
also contributes to the final results.
The $3N$-photoabsorption process must contribute the proton spectrum
around this dip momentum region.
We then check the $3N$-photoabsorption process for this experiment.

Of course the semi-classical approach cannot well describe the structure
of so small nuclei such as $^{3}$He. 
However a cross-section of each step contribution is 
almost determined by the geometrical position of nucleon.
The interaction range of the two-body collisions is about 1 - 2 fm
while the root-mean square radius of $^{3}$He is about 1.8fm.
Hence there is no serious trouble to make a rough estimation 
with the QMD approach.

In Fig. \ref{gHe280} we give a cross-section of two-protons emission
from the photoabsorption by $^3$He as a function of the undetected 
neutron momentum.
Experimental data is taken from Ref. \cite{Emura}.
In order to separate contributions of the two-protons 
$(2N(pp))$-photoabsorption 
process experimentally \cite{Emura},
they chose the events without pions at the final state 
and by a experimental trigger that $p \ge 300$MeV/c 
and $\theta = 15^{\circ} - 165^{\circ}$ for the two emitted protons.
This experimental condition selects the process
in which at least two protons are concerned with the photoabsorption.
The calculation also follows this condition.
In this figure the absolute value is arbitrarily  because the normalization of 
the experimental data has not been determined.

We again decompose the total yield to 0-step (dotted line)
and higher step contributions (dashed line) by 
the step-number of the undetected neutron.
In this case, the 0-step process (dotted line) means that the photon is 
absorbed by the two protons, and that the neutron is a spectator.
In the other case of the multistep contributions above 1-step, 
three nucleons are all related with the photoabsorption process.
Though we only treat the sequential binary process, this multistep process 
is effectively regarded as the $3N$-photoabsorption process, 
while the 0-step process is the $2N(pp)$-photoabsorption process.
In Ref. \cite{Emura}, they also decompose the total yield to the two 
contributions, one from the spectator neutron for $2N(pp)$ photoabsorption and 
the other from the emitted neutron for the $3N$-photoabsorption processes.
We also plot their decomposition of the $2N(pp)$ process (thin dashed line) 
and the $3N$ process (long dashed line) in the same figure.
 
The 0-step contribution appears in the low momentum region 
as a narrow peak.
The peak position of the 0-step process ($2N(pp)$-photoabsorption)
is lower than that from the experimental analysis.
This peak is sensitive to the momentum distribution of the neutron 
in the $^3$He, since the neutron is a spectator.
The semi-classical approach cannot describe correctly the momentum 
distribution of nucleons particularly the high momentum component
in small nuclei such as $^3$He.
This is the reason of this discrepancy.

As for the $3N$-photoabsorption process, 
we can see that the second peak is  much lower
than the first peak in our calculation, which disagrees with the experimental
result.
Namely our simulation does not make so large $3N$-photoabsorption process.
The $3N$-photoabsorption process should contribute the cross-section 
between the QF and the $2N$-photoabsorption processes.
The shortage of the $3N$-photoabsorption is the origin of the dip between 
the QF contribution and
the $2N(pp)$ contribution, seen in Fig. \ref{gC375}.

Since both the $2N$- and $3N$-photoabsorption processes make no pion events 
at the final state, 
the $\pi$-absorption must play an important role in these absorption processes.
Then we can get some information on a reason of the dip by studying 
the pion photoproduction.
In Fig. \ref{tot-pi0} we show the results of the integrated cross-sections for 
neutral pion photoproduction in the $^{12}$C, $^{27}$Al and  $^{63}$Cu 
targets, and compare them with the experimental data \cite{Arends86}.
The results of full QMD are denoted by the solid lines, 
while the long-dashed lines indicate the result without $\pi$-absorption. 
The calculated total cross-sections overestimate the data at the peak energy
of the $\Delta$ resonance and decrease faster at lower and higher energies.
This behavior is also seen in other calculations 
\cite{carrasco,hombach,Arends86}.
The overestimate of the calculation becomes larger with increasing
mass number $A$.

These results show that the $\pi$-absorption is too small in our approach.
In order to examine the effects of the $\pi$-absorption,
we make another test calculations by using a fixed pion mean-free path 
for the $\pi$-absorption process.
The pion mean-free path $\lambda$ is estimated to be 5 fm \cite{BUU4}.
>From the mean-free path  we estimate the $\pi$-absorption 
and the photoproduction cross-section in the following way.

We factorize the absorption rate 
$\exp(- <l> /{\lambda})$ to the results without the $\pi$-absorption, where
$<l>$ is a mean distance of pion propagating in nuclear medium.
Assuming hard sphere with a radius $R$ for nucleus,
we can estimate $<l>$ as
\begin{equation}
<l> = 
\frac{ \int_{\Omega} d{\bf r}_1 \int_{S} d{\bf r}_2 |{\bf r}_1 - {\bf r}_2|}
{ \int_{\Omega} d{\bf r}_1 \int_{S} d{\bf r}_2 }
= \frac{6}{5} R,
\end{equation}
where $\Omega$ and $S$ indicate the volume and surface integrals in
the hard sphere with radius $R$, respectively.
Here the pion is considered to be produced randomly at the position inside 
the target nucleus, and the $\Delta$ is not directly taken into account
in this estimation.

The results of this test calculation is shown by the dashed lines.
They nicely reproduce the experimental result around the {$\Delta$} peak, 
though it is slightly lower at lower and higher energies.
We then need another $\pi$-absorption process.
It may be a pure $2N$ $\pi$-absorption process which is not taken into account 
in our approach.
This pure $2N$ $\pi$-absorption process is given by the detailed valance 
of the s-wave pion production.
The production rate of this $s$-wave process is small, but the absorption rate
of the $2N$ process is not so small \cite{eng94}.
This fact suggests us one of the answers of the too small $3N$-photoabsorption
process in Fig. 5.
If the pion is absorbed by two nucleons, as a result, three nucleons share
the photon energy.
This is an additional $3N$-photoabsorption process.

However it is not so easy to introduce the pure $2N$ $\pi$-absorption in
the actual simulation.
We have to treat three body collisions,
which is not impossible but difficult in the actual simulation, 

Engel et al. \cite{eng94} suggested an easy method of 
the effective treatment of the $2N$ $\pi$-absorption process
by factorizing a density-dependent factor to the cross-section 
of $N + \Delta \rightarrow N + N$; later Hombach et al. 
\cite{hombach} also used the same method  in the photoreaction.
Here we test this approximate method. 
In order to estimate this effect in a rough way,
we increase this cross section by  four times; 
we call this test simulation ``QMD/T1''.
We then calculate the total $\pi^0$ production cross-sections
at the photon momentum $q = 375$MeV/c as a function of the target mass.
The results are shown in Fig. \ref{tot-A}.
The solid, dashed and broken lines indicate the results of QMD, QMD/T1, and no
absorption, respectively. 
The full square show experimental data taken from \cite{Arends86}.
The QMD/T1 slightly improves the result but still overestimates 
experimental data.

Furthermore we calculate again the proton and pion momentum-spectra 
for $\gamma (375 MeV/c) + {\rm C}$  reactions in Fig. \ref{gC375} with QMD/T1.
The results are given in Fig. \ref{gCd} by dashed lines.
The QMD/T1 slightly reduces the $\pi^0$ cross-section overall, and 
enhances the proton cross-section around $2N$-absorption energy region.
However the dip between the QF peak and the $2N$-absorption peak
exist in the QMD/T1 calculations.
The enlarged ${\sigma_{{\Delta}N \rightarrow NN}}$ mainly enhances 
the {$\Delta$} absorption process at the second step which contributes
the $2N$-photoabsorption part, but it does not contribute higher multisteps 
such as the $3N$-photoabsorption part.

When a pion propagates in a nucleus, it stays at a pure $\pi$ state much longer
than a $\Delta$ state because the $\Delta$ life-time is very short.
Then the enlarged ${\sigma_{{\Delta}N \rightarrow NN}}$ does not largely
change the $\pi$-absorption.
One may have an idea that we can increase the $\pi$-absorption by enlarging
the cross-section of ${\pi}N \rightarrow {\Delta}$ as well as that
of ${{\Delta}N \rightarrow NN}$.
This approximate method must increase the pion-absorption, but it cannot solve 
the problem that the enlarged ${\sigma_{{\Delta}N \rightarrow NN}}$ still overestimates 
the proton cross-section in the $2N$-photoabsorption energy region.
Hence the enlargement of ${\sigma_{{\Delta}N \rightarrow NN}}$ and
${\sigma_{{\pi}N \rightarrow {\Delta}}}$ together cannot explain the experimental
results of the proton emission and the pion production at the same time. 
Thus we need an additional absorption process without a $\Delta$ state,
i.e. the pure $2N$-absorption of pion.

Finally we examine effects of the {$\Delta$}-potential to observables
because the photoreactions in the {$\Delta$} energy region 
are studied for the purpose of the determination of {$\Delta$} properties 
in nuclear medium.
Along this line we calculate positive pion spectra with and without 
the {$\Delta$}-potential
in the $\gamma (213 {\rm MeV/c}) + ^{40}$Ca reaction.
In Fig. \ref{gCadel} we show our results at 
$\theta_\pi = 81^\circ, 109^\circ {\rm and} 141^\circ$;
the solid and dashed lines indicate the results with the {$\Delta$}-potential
same as nucleon one and those with no {$\Delta$}-potential, respectively.
Experimental data are taken from Ref.\cite{fissum}.

>From this figure, we see that our results of the pion cross sections
are not so largely different from experimental data 
though the peak position is slightly shifted to higher energy 
at $\theta_\pi = 81^\circ$ and $109^\circ$ when the $\Delta$-potential
is switched off.
In the case of no {$\Delta$}-potential the QF peaks are further shifted to
higher energy, but this difference is not significant; 
the {$\Delta$}-potential does not strongly affect the pion spectrum.
This result suggests that it is difficult to know in-medium {$\Delta$}
properties from only simply observing such as a pion spectrum.
We should investigate more complicated coincident observables.

\section{Summary}\label{summary}

In this paper we have calculated the emitted proton and produced pion 
cross-sections in the photoreaction within the framework of the QMD approach, 
and compared them with experimental data.
We focus on the examination of applying the QMD approach to 
the photoreaction, and use rather rough initial photoabsorption 
in the present analysis.
Through this work, however, we can analyze the multistep contribution and 
check the ${\pi}$-${\Delta}$ dynamics.

As for the proton spectrum, the multistep contributions are not negligibly 
small even in the forward direction where the contribution from 
the QF process is dominate.
As increasing the emission angle, the multistep contribution becomes larger,
and overlaps with the QF contribution.
It is found that the high momentum parts in the proton 
spectrum comes mainly from by the 2-step process without pion, 
which is effectively identical to the $2N$-photoabsorption process.
Thus it is very important to take account of the multistep contribution 
when comparing a theoretical result with experimental data.

We have examined how the {$\Delta$}-potential affects 
the pion spectrum.
We cannot find any significant difference between two kinds of results of 
the inclusive pion spectra with and without 
the $\Delta$-potential.

The cross-section around QF peak of proton are overestimated at 
$\theta_{p} = 30^{\circ}$ and underestimated at $\theta_{p} = 52^{\circ}$.
The discrepancy of the cross-section around the QF-peak may be improved 
by using the realistic angular distribution of ${\pi}$-production 
at the initial channels.
In order to include it, we have to introduce the anisotropic
decay of {$\Delta$}.
It is, however, not easy because the spin degree is not involved
in our semi-classical approach.
We then need to consider a technical extension of our approach such as 
the p-wave decay of {$\Delta$} given by  Engel et al. \cite{eng94}.

Furthermore we have the other problems: too large pion-photoproduction
and a dip between the QF-peak and the $2N$-absorption peak,
which is not observed in the experimental data.
These two problems are supposed to be caused by the same reason, i.e.
lack of the pure $2N$ $\pi$-absorption 
process which dose not involve the intermediate {$\Delta$}.
It is shown that the artificial enhancement of the $N{\Delta} \rightarrow NN$ 
cross-section (QMD/T1 calculation) does not give sufficient absorption of pion.
>From both results of the proton emission and the pion production, thus,
we can conclude that one should treat directly 
the pure $2N$ $\pi$-absorption process, i.e. three body collisional process,
in the dynamical simulation.

In future we will improve our approach by introducing the pure $2N$-absorption
process of pions and more realistic elementary photoabsorption processes
of the pion production \cite{nozawa} and the quasideutron processes
\cite{QD}.
By such improved model, we would like to investigate the proton- and 
pion-induced reactions as well as the photoreactions simultaneously.
Such works will give useful information to extend a simulation
for investigation of general reactions.

\acknowledgements

Th authors wish to thank  Drs. T. Suda and K. Maruyama
for valuable discussions and comments.


\pagestyle{empty}


~~
\begin{figure}
\noindent
\caption{The cross-section of $\gamma + p \rightarrow p + \pi^0$ (a) and  
$\gamma + p \rightarrow n + \pi^{+}$ (b) processes.
The long dashed, dashed and thin solid lines indicate contributions
of the $\Delta$ resonance, $N^{*}(1440)$ and the Born terms.
Experimental data are taken from Refs. [15,16].
}
\label{gnpi}
\end{figure}

\begin{figure}
\noindent
\caption{Proton and $\pi^0$ cross-sections at laboratory angle 
$\theta = 30^{\circ}$, $60^{\circ}$ and $90^{\circ}$ 
for 375 MeV photon + $^{12}$C reactions.
The left columns show the proton spectra, and   
the right ones show the $\pi^0$ spectra
The total (thick solid line), 1-step (dashed line), 2-step (broken line) 
and 3-step (thin solid line)
QMD cross-sections are compared with experimental data [17].}  
\label{gC375}
\end{figure}

\begin{figure}
\noindent
\caption{Proton momentum-spectrum at laboratory angle $\theta = 30^{\circ}$ 
for 375 MeV photon + $^{12}$C reaction.
The upper figure is same as the first figure in Fig. 2.
In the lower figure the dashed and broken lines indicate contribution
from events with zero and one pion at final state.
The chain-dotted line shows the total result with all Born initial state.}  
\label{gC375ch}
\end{figure}

\begin{figure}
\noindent
\caption{Proton energy-spectrum at laboratory angle $\theta = 51^{\circ}$ 
for 390 MeV/c photon + $^{12}$C and $^{48}Ti$ reactions.
Experimental data are taken from Refs. [21].}
\label{g390p}
\end{figure}

\begin{figure}
\noindent
\caption{
Neutron momentum distribution for 280 MeV/c photon + {$^3$He} reactions.
The condition is given in the text. 
The thick solid line shows the total QMD results, and  
the dotted and dashed lines denote contributions
from 0-step and higher steps, respectively.
The experimental data  are taken from Ref. [5]
The thin dashed and broken lines indicate $2N(pp)$ and $3N$ processes,
respectively, which are given from experimental analysis [5].
} 
\label{gHe280}
\end{figure}

\begin{figure}
\noindent
\caption{Total $\pi^{0}$ cross-sections of the targets
$^{12}$C, $^{27}$Al and $^{63}$Cu.
QMD cross-sections are compared with experimental data [22].}
\label{tot-pi0}
\end{figure}

\begin{figure}
\noindent
\caption{Target mass-number dependence of the total $\pi^{0}$ cross-sections 
at $q = 375$MeV/c.
The solid, dashed and broken lines indicate the results of QMD, QMD/T1, and no
absorption. 
The full square show experimental data taken from Ref. [22] }
\label{tot-A}
\end{figure}

\begin{figure}
\noindent
\caption{Proton and $\pi^0$ cross-sections at laboratory angle 
$\theta = 30^{\circ}$, $60^{\circ}$ and $90^{\circ}$ 
for 375 MeV/c photon + $^{12}$C reactions.
The left columns show the proton spectra, and   
the right ones show the $\pi^0$ spectra
The solid and dashed lines indicate the results by QMD and QMD/T1, respectively.
}  
\label{gCd}
\end{figure}

\begin{figure}
\noindent
\caption{$\pi^{+}$ cross-sections at laboratory angle 
$\theta = 81^{\circ}$, $109^{\circ}$ and $141^{\circ}$ 
for 213 MeV/c photon + $^{40}$Ca reactions.
The solid line shows the QMD results with the {$\Delta$} mean-field
same as that of nucleons. 
The dashed line shows the results with the no {$\Delta$} mean-field.
The experimental data are taken from Ref. [25]}  
\label{gCadel}
\end{figure}

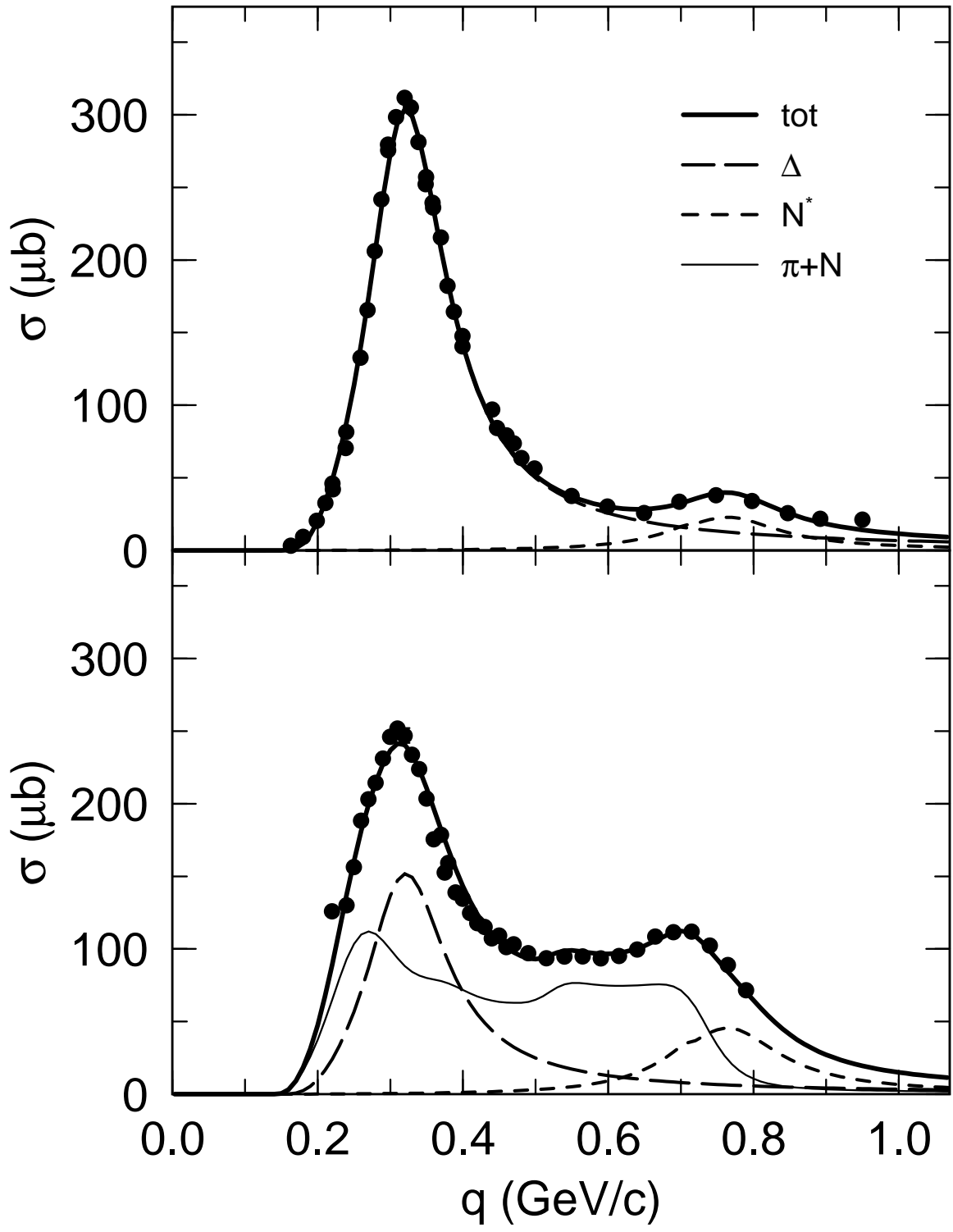
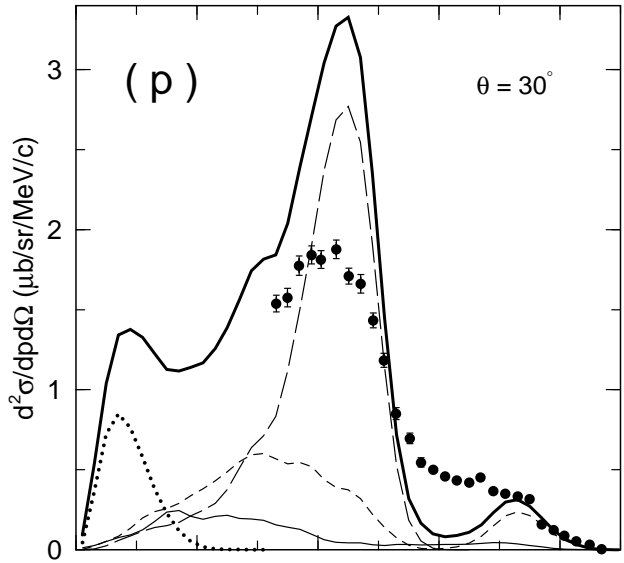
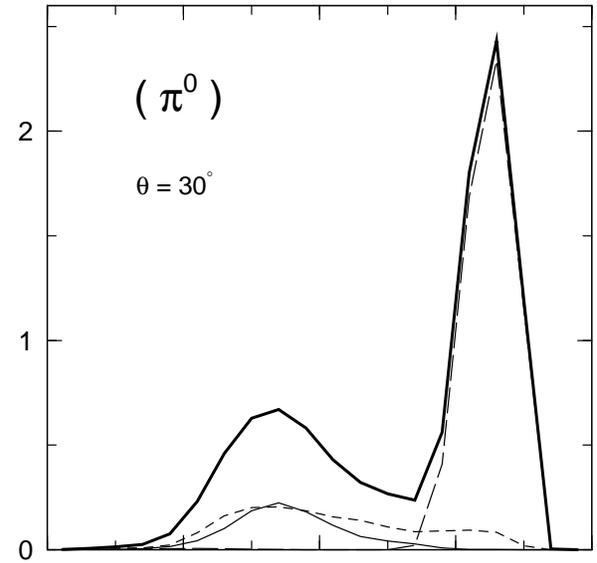
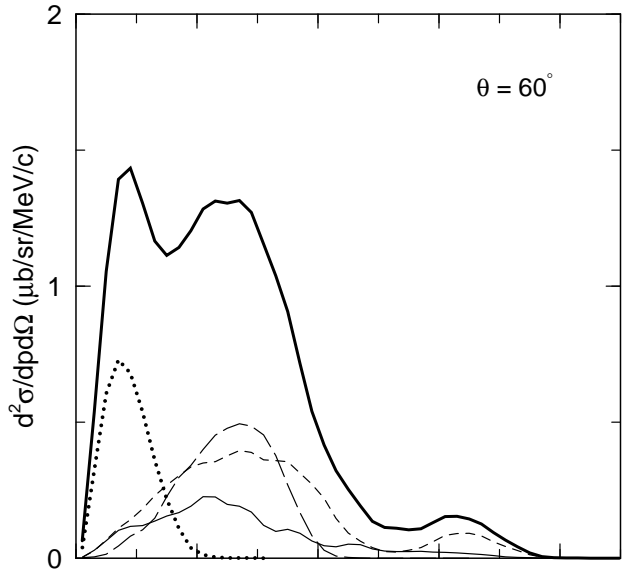
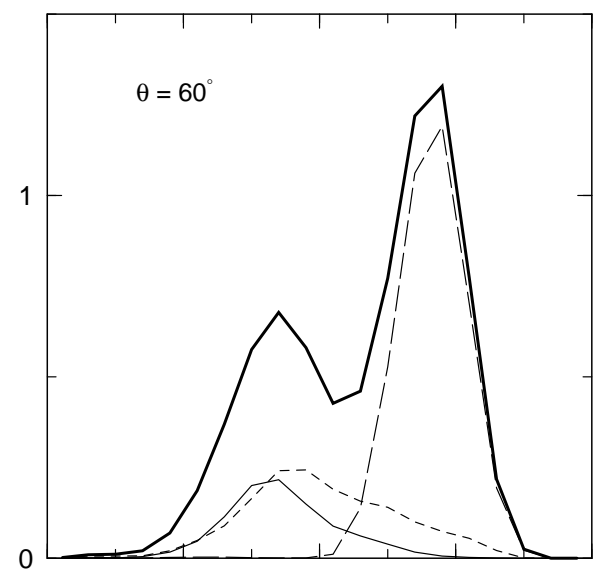
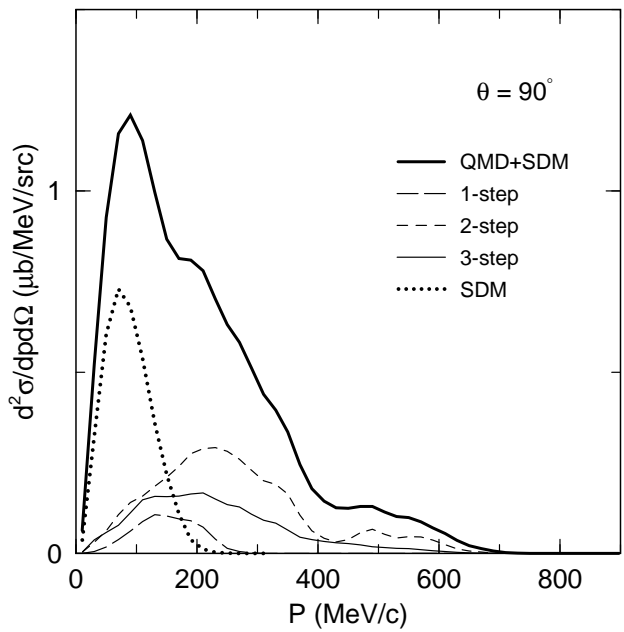
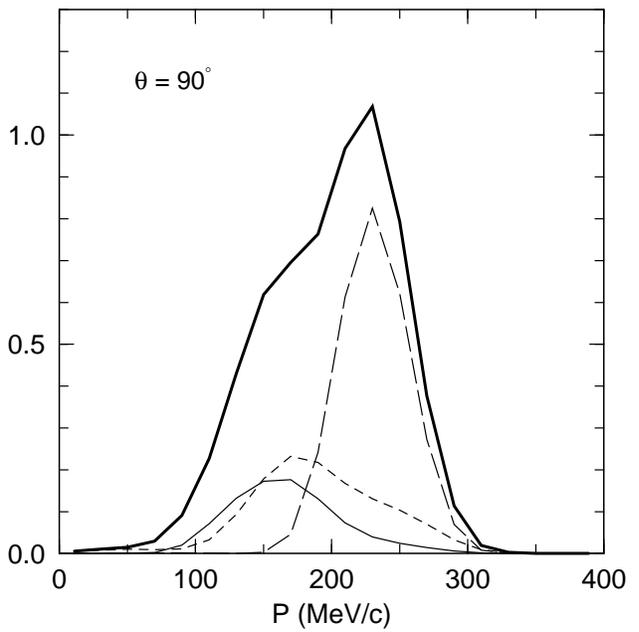
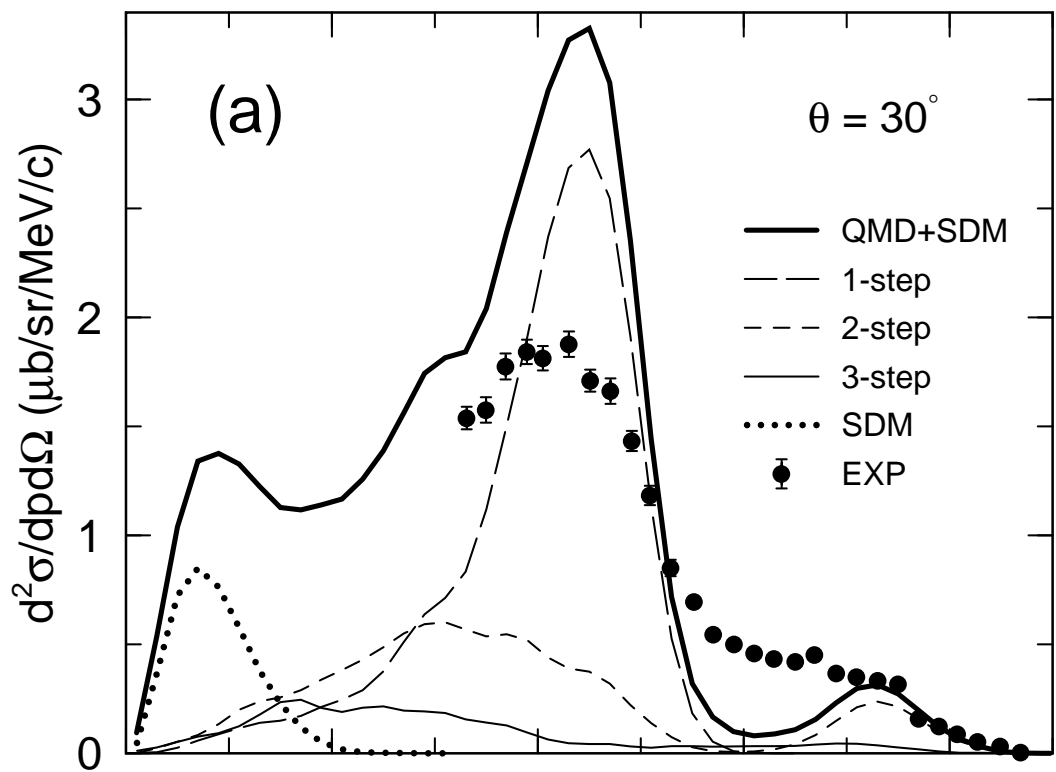
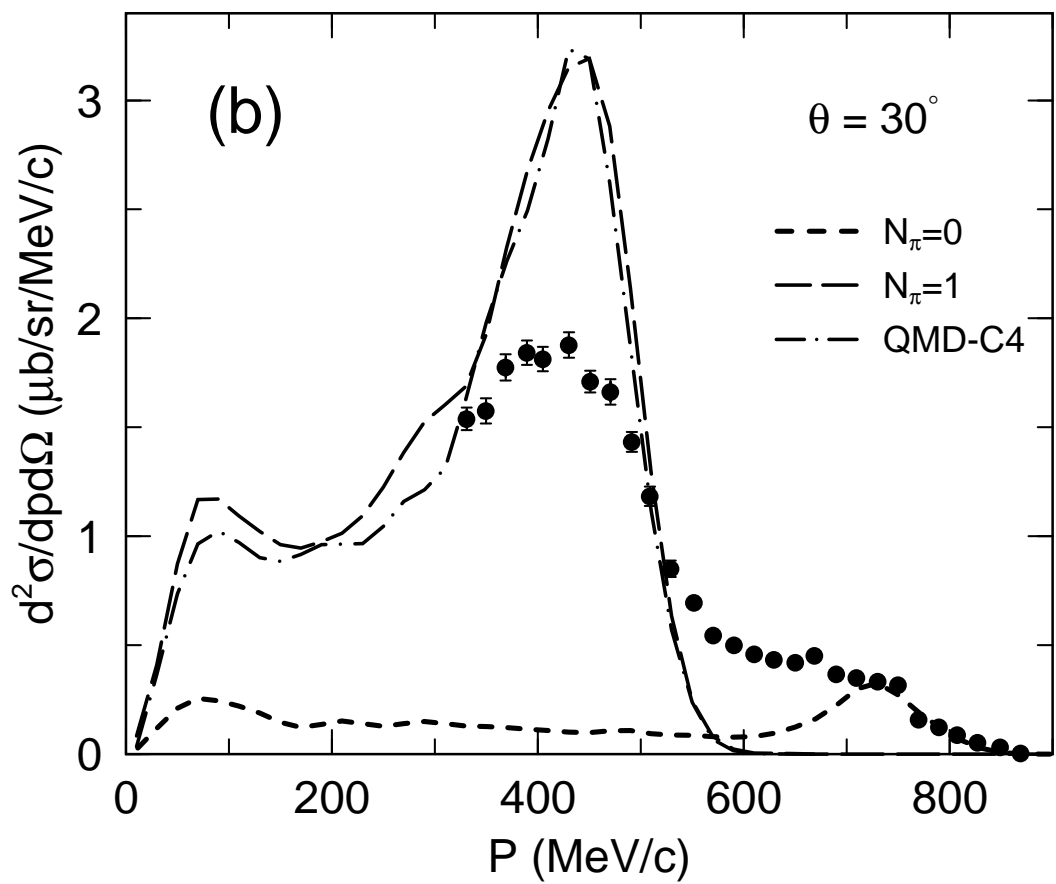
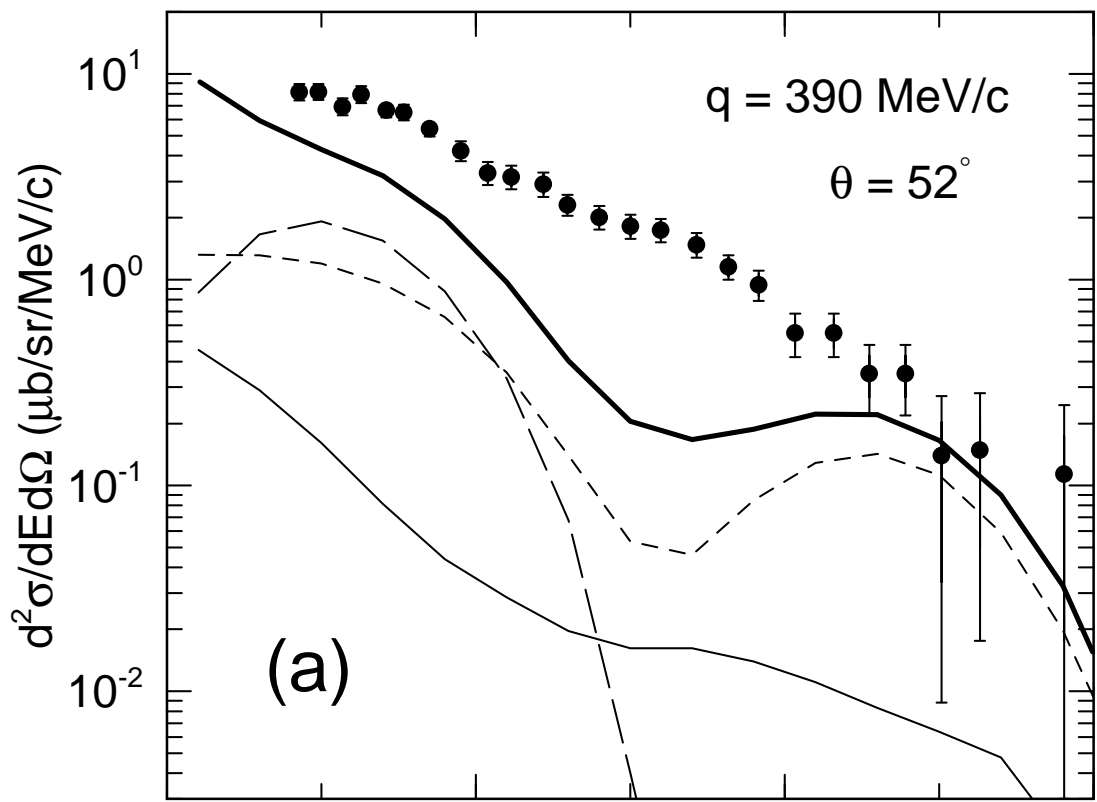
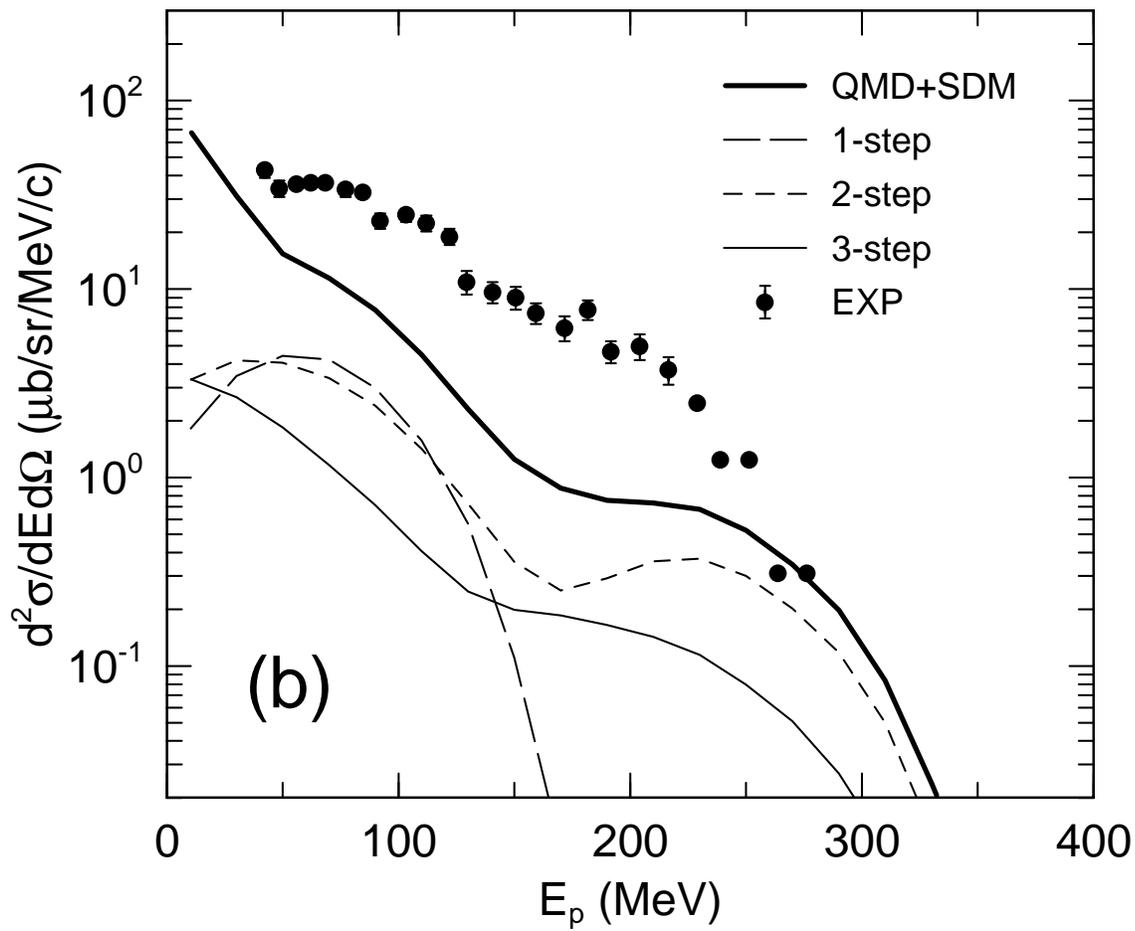
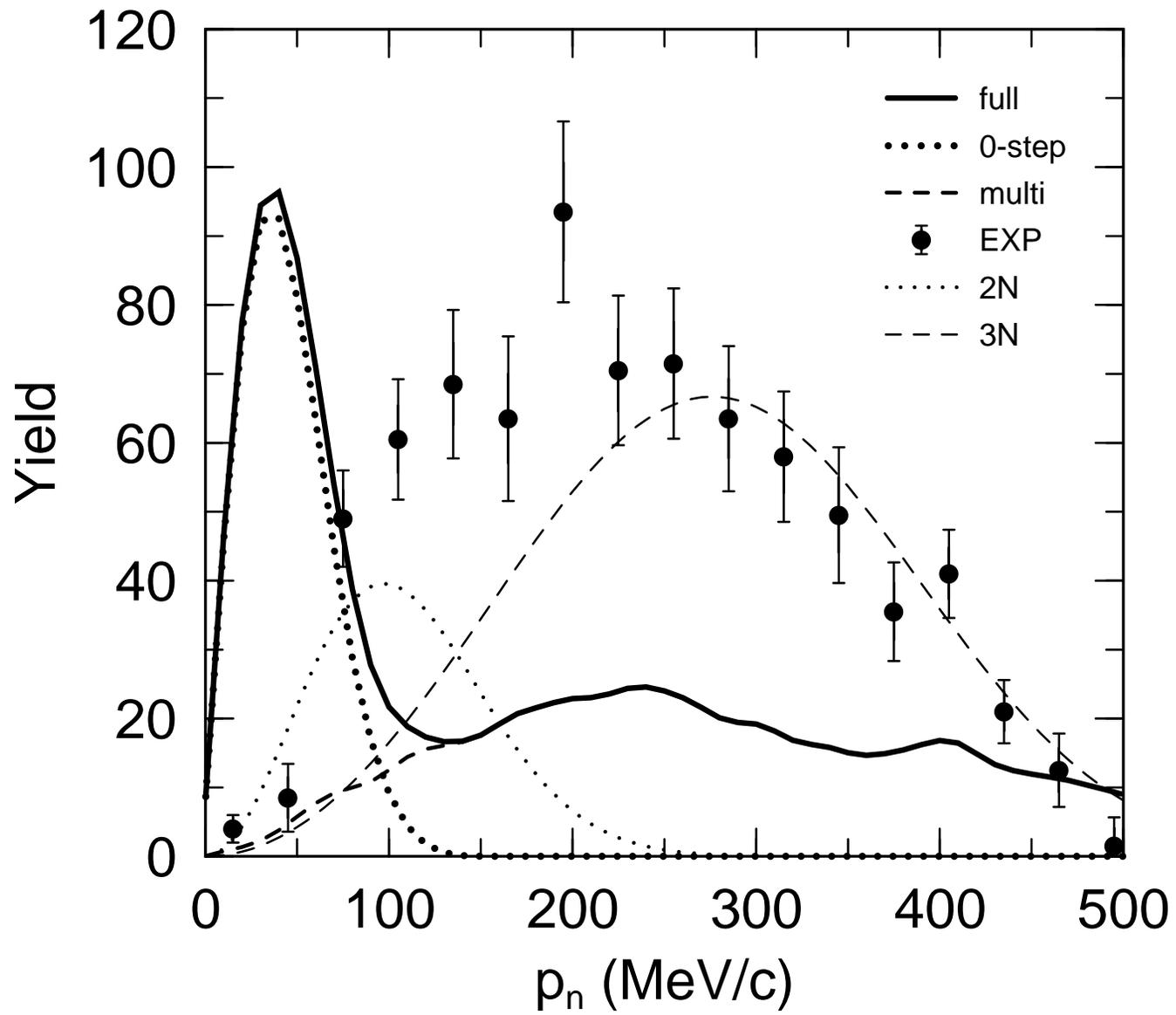
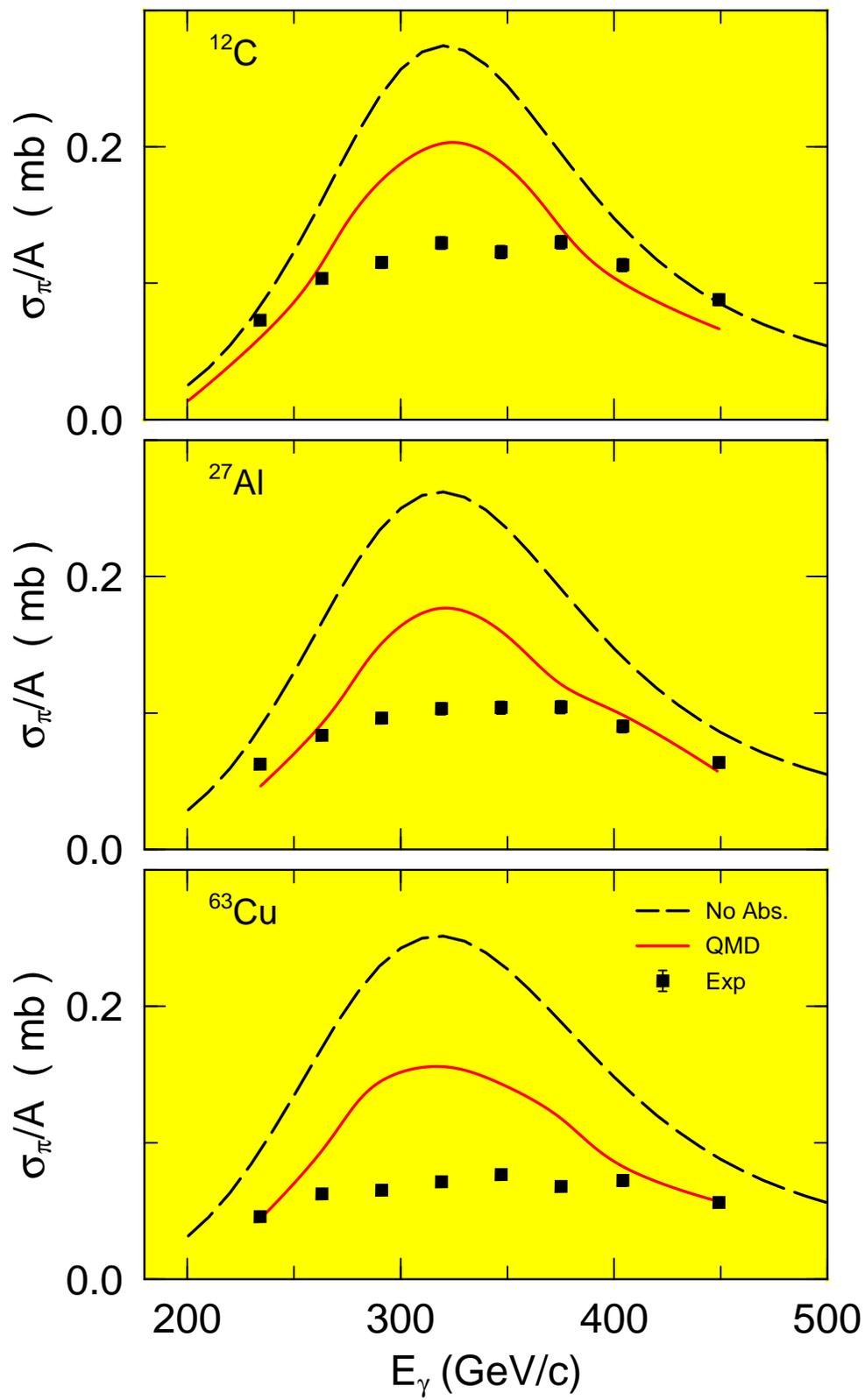
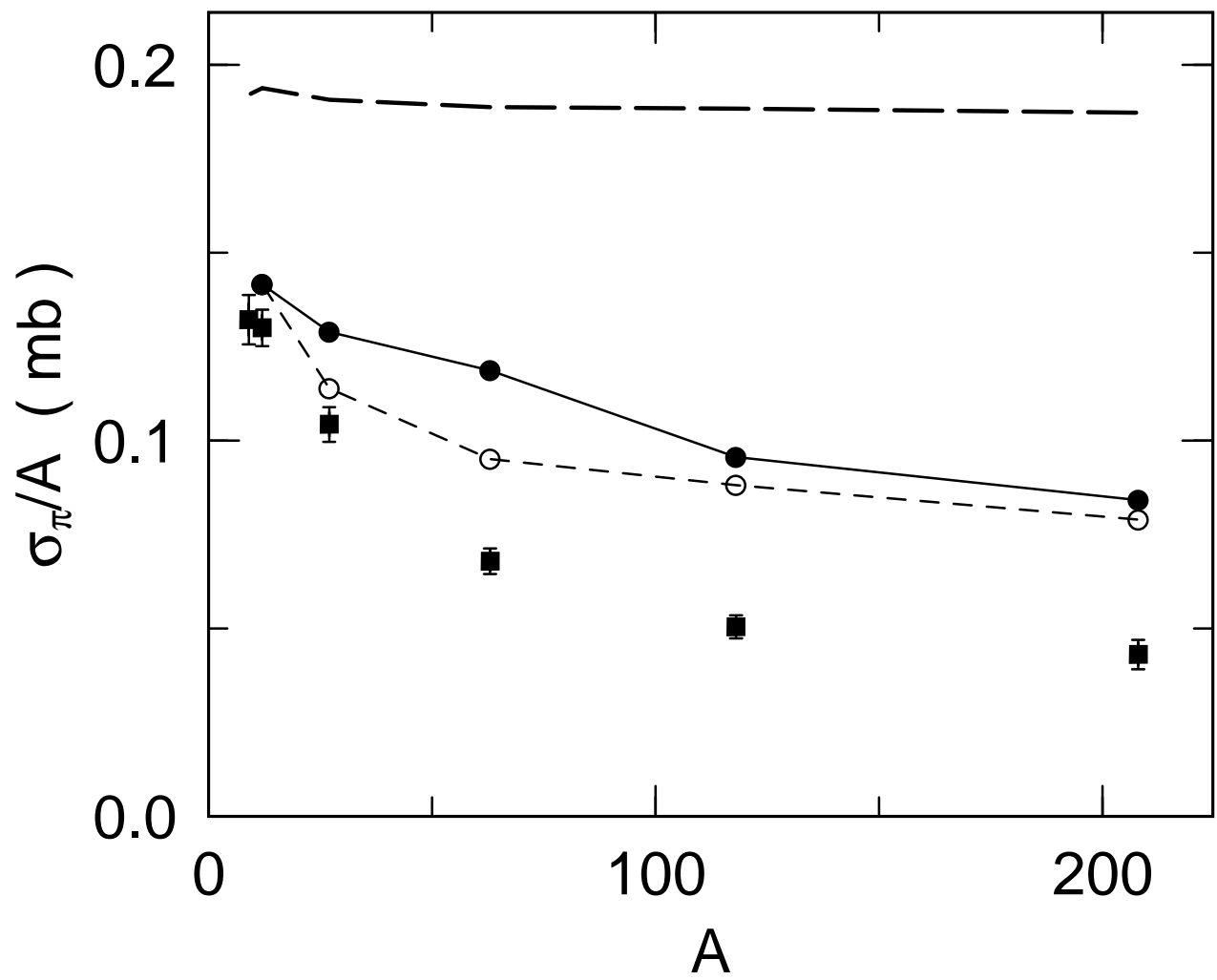
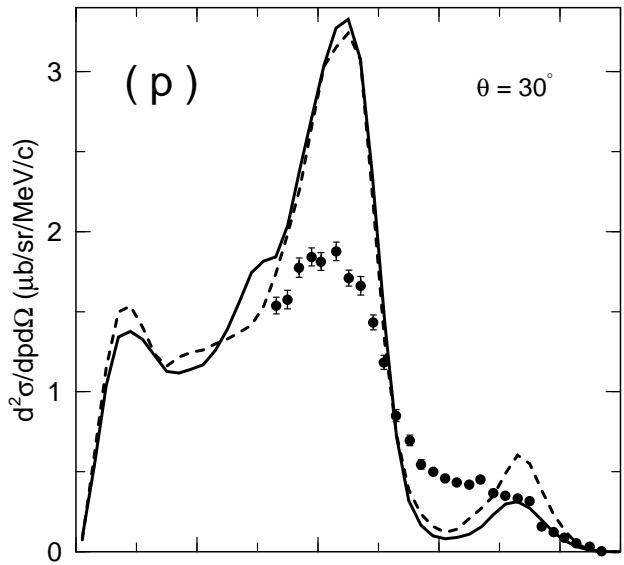
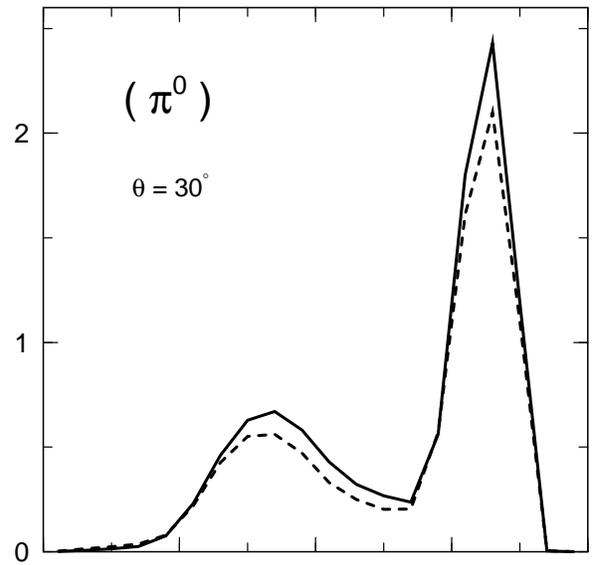
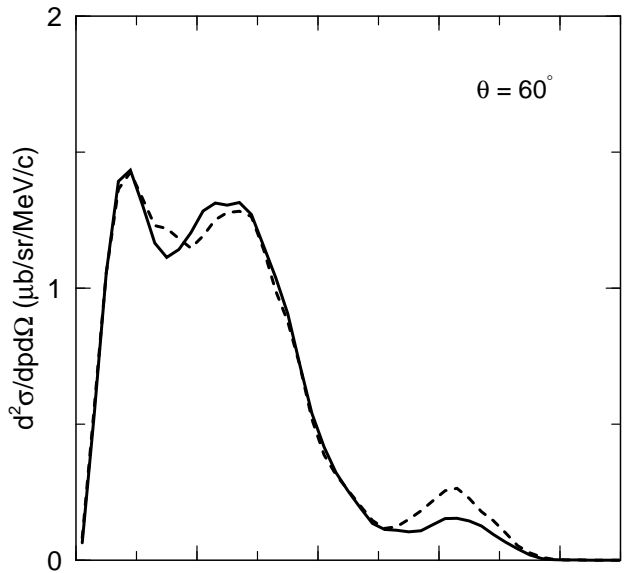
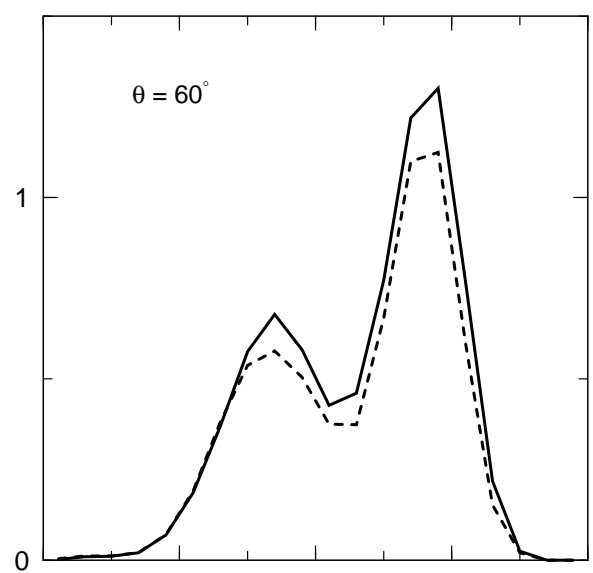
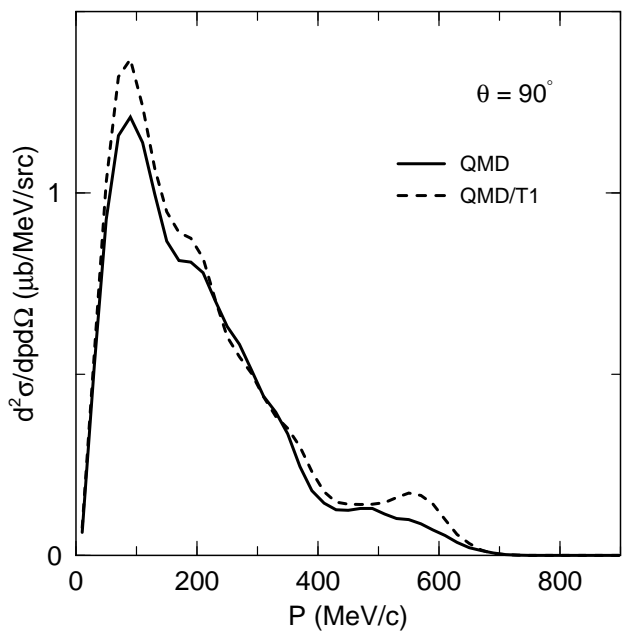
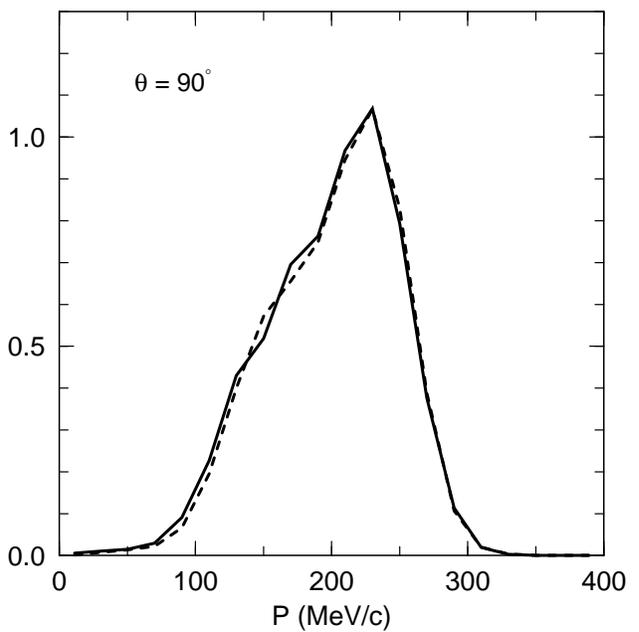
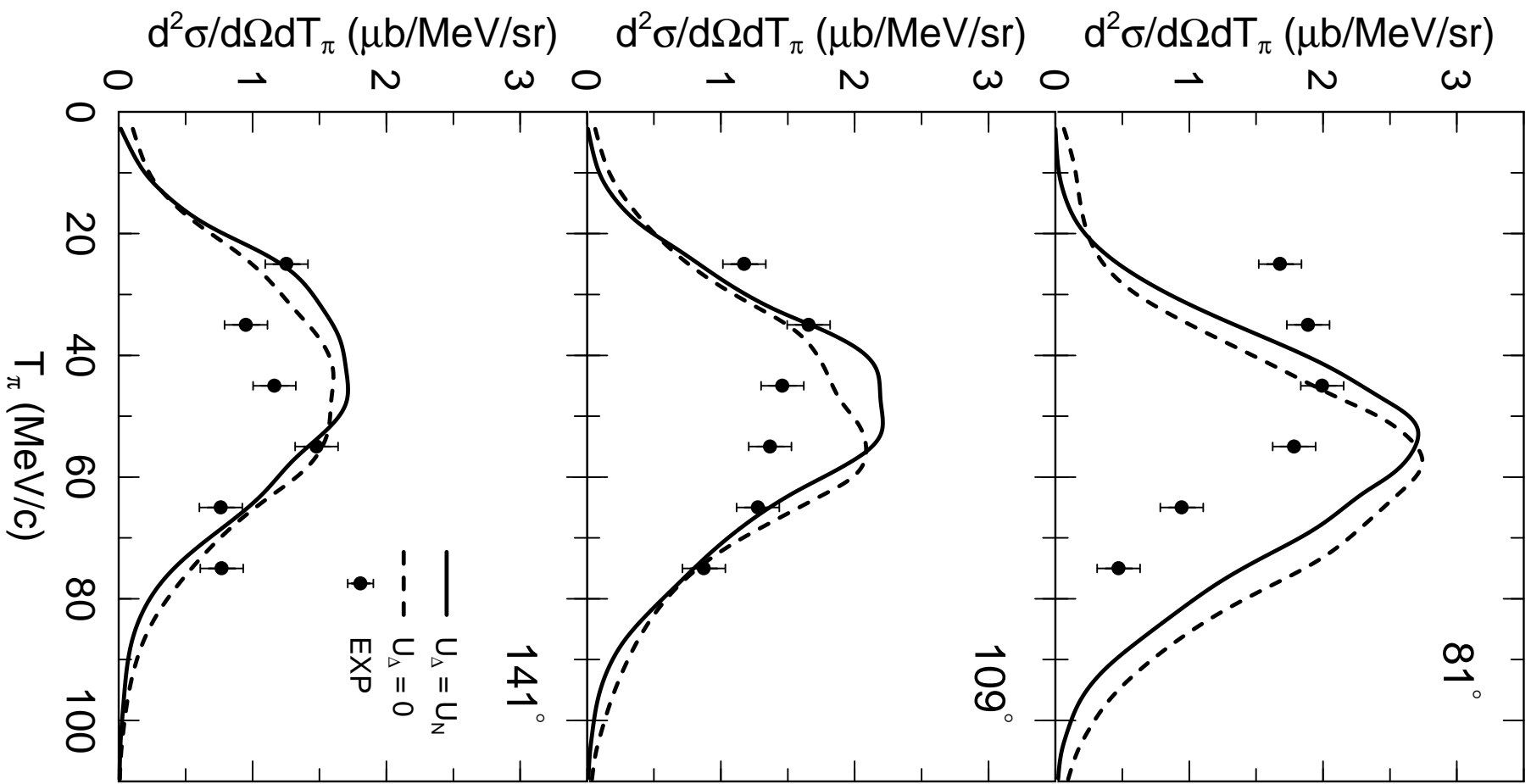
\end{document}